\documentclass[prb,preprint,showpacs,preprintnumbers,amsmath,amssymb]{revtex4}

\usepackage{graphicx}
\usepackage{dcolumn}
\usepackage{bm}

\begin{document}


\title{Pressure dependence of the  superconducting transition and  electron correlations in  Na$_{x}$CoO$_{2}$$\cdot 1.3$H$_{2}$O }

\author{E.~Kusano$^1$}%
\author{S.~Kawasaki$^1$}%
\author{K.~Matano$^1$}%
\author{Guo-qing~Zheng$^1$}%
\author{R.L.~Meng $^2$}%
\author{J.~Cmaidalka $^2$}%
\author{C.W.~Chu $^{2,3,4}$}%

\address{$^1$ Department of Physics, Okayama University,  Okayama 700-8530, Japan}
\address{$^2$ Department of Physics and TCSAM, University of Houston, TX 77204-5932, USA.}
\address{$^3$ Lawrence Berkeley National Laboratory, 1 Cyclotron Road, Berkeley, CA 94720, USA} 
\address{$^4$ Hong Kong University of Science and Technology, Hong Kong, China}

\date{21 June, 2007}

\begin{abstract}
We report $T_c$ and  $^{59}$Co nuclear quadrupole resonance (NQR) measurements  on the  cobalt oxide superconductor Na$_{x}$CoO$_{2}$$\cdot 1.3$H$_{2}$O ($T_c$=4.8 K) under hydrostatic pressure ($P$) up to 2.36 GPa.   $T_c$   decreases with increasing pressure at an average rate of -0.49$\pm$0.09 K/GPa. 
 At low pressures $P\leq$0.49 GPa, the decrease of $T_c$ is accompanied  by a weakening of 
the  spin correlations   at a finite wave vector and a reduction of the density of states (DOS) at the Fermi level. At high pressures above 1.93 GPa, however, the decrease of $T_c$ is mainly due to a reduction of the DOS.   These results indicate that the electronic/magnetic state of Co is primarily responsible for the superconductivity. 
  The spin-lattice relaxation rate $1/T_1$ at  $P$=0.49 GPa shows a $T^3$ variation below $T_c$ down to $T\sim 0.12T_c$, which provides compelling evidence for   the presence of  line nodes in the superconducting  gap function.

\end{abstract}


\hspace{5cm} Phys. Rev.B {\bf 76}, 100506(R) (2007)

\maketitle

\sloppy

The recently discovered superconductivity in cobalt oxide Na$_{x}$CoO$_{2}\cdot$1.3H$_{2}$O has attracted considerable attention \cite{Takada}.
This compound bears some similarities to the high-$T_{c}$ copper oxide superconductors, although Co forms triangular lattice instead of square lattice in the cuprates.
In an early study on a sample with large Na-content, $x$=0.31 and $T_c$=3.7 K, it has been found that the spin-lattice relaxation rate $1/T_1$ shows a $T^{n}$ decrease with no coherence peak just below $T_c$, which suggests non-$s$ wave superconducting state  \cite{Fujimoto}. However, the crossover from  $n$=3 to $n$=1 at low temperatures, which was subsequently confirmed by other groups, has generated debates on whether there are nodes in the gap function \cite{Balatsky,Wang}.  Later  measurements on a low-$x$ sample with  high $T_c$ ($x$=0.26, $T_c$=4.6 K) reveal that $1/T_1$ follows a $T^{3}$ variation all way down to $T=T_c/6$, indicating unambiguously that there exist line nodes in the gap function \cite{Zheng}.     It is also worth noting that the spin fluctuations at finite wave vector, most likely of antiferromagnetic (AF) origin,  increase with decreasing Na-content, and become strongest at $x \sim$ 0.26 where $T_c$ is the highest \cite{Zheng}.

To investigate the spin-pairing symmetry, the Knight shift has been measured both in a powder sample \cite {Kobayashi} and in a single crystal \cite{Zheng2}. It has been found that the shift decreases below $T_c$ along both the $a$-axis and $c$-axis directions \cite{Zheng2}. This indicates that the Cooper pairs are in the singlet form. Therefore, the experimental results strongly suggest that the superconductivity is of $d$-wave symmetry. 

 In this paper, we report NQR (nuclear quadrupole resonance) studies on Na$_{x}$CoO$_{2}$$\cdot 1.3$H$_{2}$O with the highest $T_c$=4.8 K under pressure ($P$), in order to get a hint as to the mechanism of the superconductivity. We investigate the correlation between $T_c$ and the electronic states.  We have measured $T_c$, NQR frequency $\nu_Q$, and the spin-lattice relaxation rate $1/T_1$ under pressures up to 2.36 GPa. We find that $T_c$ decreases with increasing pressure at an average rate of -0.49$\pm$0.09 K/GPa. At low pressures $P\leq$0.49 GPa, the decrease of $T_c$ is  due to a weakening of the  spin correlations    and a reduction of the density of states (DOS) at the Fermi level. At high pressures above 1.93 GPa, in contrast, the decrease of $T_c$ is due mainly to a reduction of the DOS .  Our results indicate that the magnetic/electronic state of Co is primarily responsible for the superconductivity. Also, the spin-lattice relaxation rate $1/T_1$ at $P$=0.49 GPa shows a $T^3$ variation below $T_c$ down to $T\sim 0.12T_c$, as at  ambient pressure \cite{Zheng}, which adds additional evidence for   the presence of  line nodes in the superconducting  gap function.

The Na$_x$CoO$_2$$\cdot$yH$_2$O powder \cite{Lorenz} was synthesized following Ref. \cite{Takada}. 
 The Na content of this sample was not determined independently. Comparison of  the   $c$-axis length, $\nu_Q$ and $1/T_1$ data with the previous group of samples \cite{Zheng} shows that this sample    has a  $x$=0.26$\sim$0.28. 
$T_c$ was determined from the ac susceptibility and $1/T_1$. 
$^{59}$Co NQR measurements were carried out by using a phase-coherent spectrometer. 
 The nuclear magnetization decay curve is excellently fitted to the theoretical formula \cite{Mac}, 
\begin{eqnarray}
\frac{M_0-M(t)}{M_0}=0.095exp(-3t/T_1)+0.095exp(-9.5t/T_1)+0.819exp(-19t/T_1),
\end{eqnarray}
 with a unique $T_1$ component.
The hydrostatic pressure was applied by utilizing a NiCrAl/BeCu piston-cylinder type cell, filled with Daphne 7373 oil as a pressure-transmitting medium\cite{Andrei}. The value of pressure at low temperatures was determined from the  $T_c$ value of Sn metal  measured by a conventional four terminal method \cite{Chu}. For $P$ other than 0.49 GPa, all measurements were done during the same experimental run at each pressure. 
For $P$=0.49 GPa, measurements above 1.4 K were carried out during one experimental run, then the pressure cell was warmed up and attached to  a  $^3$He refrigerator for data acquisition below 1.4 K.  

Figure 1 shows the ac susceptibility measured using the {\it in situ} NQR coil. $T_c$ was determined as the onset temperature of the diamagnetism, which is in agreement, within $\pm$0.1 K,  with the temperature at which $1/T_1$ decreases (see below). 

\begin{figure}
\begin{center}
\includegraphics[scale=0.5]{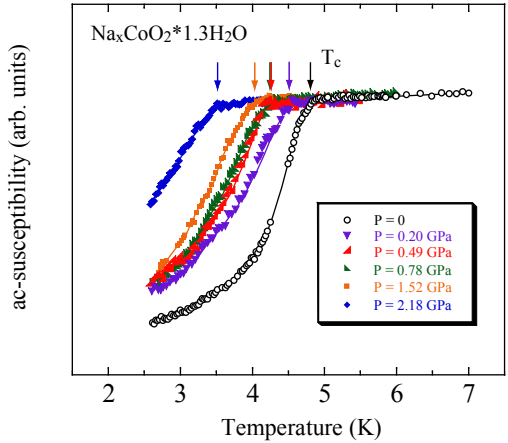}
\caption{ (Color online) Typical examples of the ac susceptibility measured using the NQR coil. The arrows indicate $T_c$ at different pressures. }
\label{fig:1}
\end{center}
\end{figure}

 Figure 2 shows  the $\pm$ 5/2$\leftrightarrow$$\pm$7/2 NQR transition line at three typical  pressures. For convenience, we call this resonance frequency 3$\nu_Q$, which  increases with increasing  pressure. It is noted that the width of the transition does not depend on pressure, indicating that the quality of the sample is unaffected by applying pressure.


\begin{figure}
\begin{center}
\includegraphics[scale=0.5]{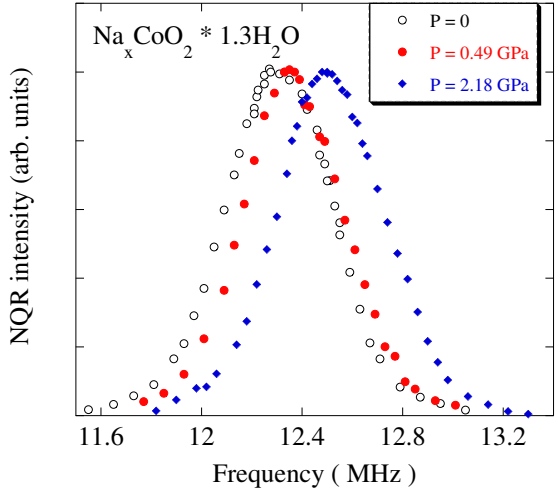}
\caption{ (Color online) $^{59}$Co  NQR spectra at the $\pm$ 5/2$\leftrightarrow$$\pm$7/2 transition under three different pressures. }
\label{fig:1}
\end{center}
\end{figure}

Figure 3 shows the temperature dependence of the quantity $1/T_1T$ at ambient and high pressures. The data for pressures of 1.27 and 1.93 GPa are very similar to those at 0.49 GPa and are not shown for the sake of clarity. Overall,  the absolute value of $1/T_1T$ decreases with increasing $P$. Also, the extent that  $1/T_1T$ increases progressively upon lowering $T$  changes as well (see later discussion). It has been noted that the non-Korringa behavior, namely,  the increase of $1/T_1T$  upon lowering $T$, is a signature of spin correlations \cite{Fujimoto,Zheng}. Since the Knight shift along both $a$- and $c-$axis decreases with decreasing $T$ and becomes a constant at $T_c \leq T \leq$ 100 K, the spin correlations are not ferromagnetic in origin \cite{Singh}, but rather point to antiferromagnetic spin fluctuations.

\begin{figure}
\begin{center}
\includegraphics[scale=0.5]{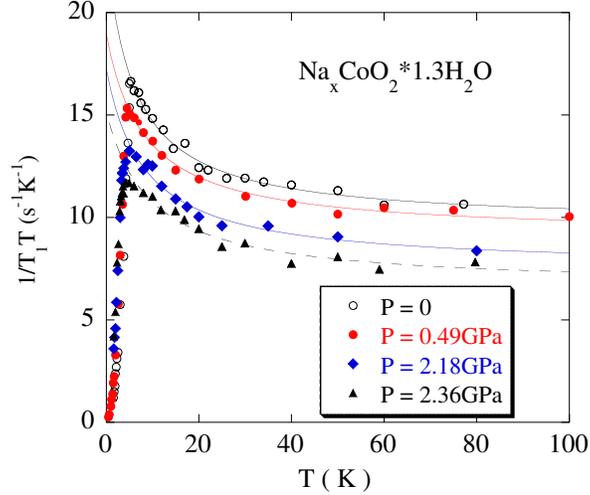}
\caption{(Color online) The quantity $1/T_1T$ as a function of temperature at ambient and high pressures.   The curves are fits to $1/T_1T = (1/T_1T)_0$+ 80 [s$^{-1}$]/$(T+\theta)$. The resulting  $(1/T_1T)_0$ and $\theta$ are plotted in Fig. 4. }
\label{fig:2}
\end{center}
\end{figure}

In order to quantify the discussion,  we analyze our data in the same way as for the ambient pressure data.  In a general form, $1/T_1T$ is written as 
\begin{eqnarray}
1/T_1T=\sum_q A_q^2 \chi^{,,} (q,\omega)/\omega
\end{eqnarray}
 in the $\omega \rightarrow$0 limit, where $A_q$ is the hyperfine coupling constant, and $\chi^{,,} (q,\omega)$ is the imaginary part of the $q$-dependent, dynamical susceptibility. 
If one assumes that there is a peak around the AF wave vector $Q$, then one may have the following approximation: 
\begin{eqnarray}
1/T_1T =(1/T_1T)_{AF}+(1/T_1T)_{0}
\end{eqnarray}
where $(1/T_1T)_0$ denotes the contribution from wave vectors other than $Q$, which usually is dominated by the uniform susceptibility $\chi_0$. For the AF contribution, we use the formula for a two-dimensional  antiferromagnetically-correlated metal, namely, ($1/T_1T)_{AF}=\frac{C}{T+\theta}$ (Ref. \cite{Moriya}). We find that such model fits the data quite well, as depicted by the curves in Fig. 3. The resulting fitting parameters are shown in Fig. 4. Note that a small value of $\theta$ means that a system is close to the AF instability ($\theta$=0). Therefore, the increase of $\theta$ to a larger positive value indicates a weakening of the antiferromagnetic correlations.
On the other hand, the decrease of $(1/T_1T)_0$  can be attributed to the decrease of density of state (DOS) at the Fermi level, $N(E_F)$, since $(1/T_1T)_0\propto N(E_F)^2$.

\begin{figure}
\begin{center}
\includegraphics[scale=0.45]{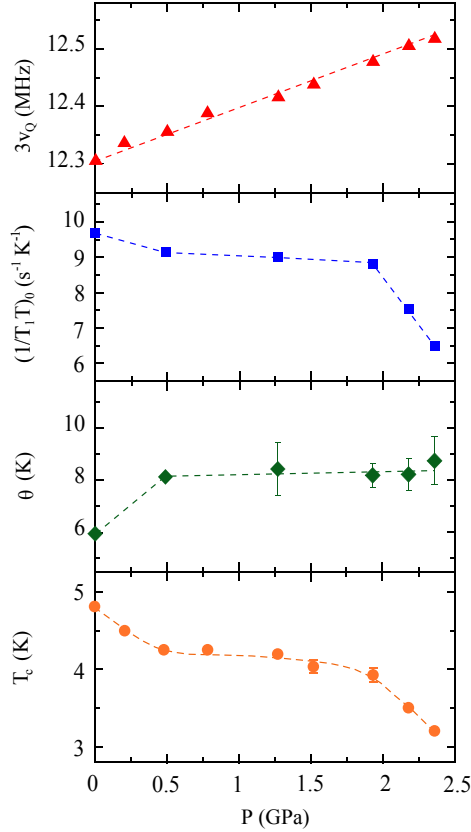}
\caption{(Color online) Various parameters obtained in this study plotted as a function of pressure. The dotted curves are guides to the eyes. The error in estimating the peak frequency 3$\nu_Q$ and $(1/T_1T)_0$ is within the size of the symbols. Note that the density of states is proportional to $\sqrt{(1/T_1T)_0}$.}
\label{fig:3}
\end{center}
\end{figure}

As seen in Fig. 4, 
 overall, $T_c$ decreases with increasing pressure, but non-monotonically. The average rate of $\frac{dT_c}{dP}$ in the whole $P$ range is -0.49$\pm$0.09 K/GPa. The average rate below 1.6 GPa is -0.28$\pm$0.05 K GPa, which is in good agreement with    the  rate  -0.27$\pm$0.03 K/GPa reported by Lorenz {\it et al} \cite{Lorenz} who measured the ac susceptibility in this pressure range. However, In the measurements of Lorenz {\it et al}, $T_c$ is $P$-insensitive in the range of $P\leq$0.5 GPa, while in the present study, $T_c$ is $P$-insensitive in the range of 0.49 GPa$\leq P\leq$1.93 GPa. The reason for the discrepancy is unknown at this point. 
It is noted that  $\nu_Q$ increases smoothly in the whole pressure range, which assures that the pressure is indeed transmitted to the sample.  



 Below 0.49 GPa, $\theta$ increases and  $(1/T_1T)_{0}$ decreases, which indicates that the decrease of $T_c$ is accompanied by both the weakening of the spin corelations and the reduction of DOS. The pressure effects on $T_c$, $\theta$, DOS and $\nu_Q$ are equivalent to increasing the Na content \cite{Zheng}. This suggests that the pressure effect in this range is to induce a charge re-distribution between Co and oxygen. 
On the other hand,  at pressures above 1.93 GPa, the DOS  decreases with increasing $P$  while $\theta$  is less $P$-dependent, suggesting that the reduction of $T_c$ at high pressures is primarily due to the reduction of the DOS. 
Although several mechanisms for the superconductivity \cite{Kontani,Lee,Johanes,Liu,Kuroki,Yanase}, including electron-lattice interaction \cite{Kontani}, charge order-driven mechanism \cite{Lee}, have been proposed,
our result indicates that the Co electronic state (namely, the DOS) and the magnetic fluctuations of likely antiferromagnetic origin are primarily responsible for the superconductivity.
 
 \begin{figure}
\begin{center}
\includegraphics[scale=0.5]{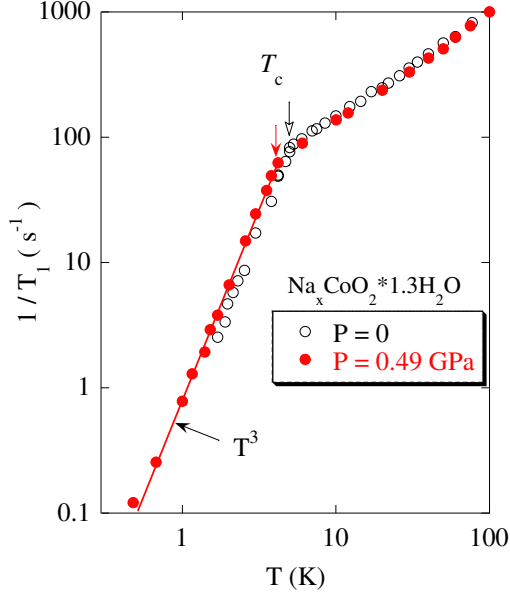}
\caption{(Color online) Temperature dependence of $1/T_1$ at ambient pressure and $P=0.49$ GPa. The arrow indicates $T_c$ at different pressures. The straight line depicts the $T^3$ variation.}
\label{fig:4}
\end{center}
\end{figure}

Finally, we turn to the superconducting state. 
 Figure 5 shows  the temperature dependence of $1/T_1$ at $P$=0 and 0.49 GPa. It can be seen that $1/T_1$ follows a $T^3$ variation down to the lowest temperature measured ($\sim 0.12T_c$). This result establishes that the sample  is free of impurity scattering, and
indicates unambiguously that there exists line-nodes in the gap function.
As we noted in  previous publications \cite{Fujimoto,Zheng}, the gap function with line nodes generates    an energy ($E$)-linear DOS at low $E$ which results in a $T^n$ ($n$=3) dependence of $1/T_1$.  Low-$E$ excitations arising from impurity scattering or other competing orders usually modify the DOS near the Fermi level,  leading to an exponent $n$ less than 3. Also, an anisotropic $s$-wave gap would give rise to a coherence peak just below $T_c$, which is incompatible with the experimental result.
It is noted that the existence of nodes is consistent with superconductivity being induced by electron correlations.

In conclusion, we have presented  systematic  $^{59}$Co-NQR measurements on the  cobalt oxide superconductor Na$_{x}$CoO$_{2}$$\cdot 1.3$H$_{2}$O under high pressures up to 2.36 GPa. We find that $T_c$ decreases with increasing pressure at an average rate of -0.49$\pm$0.09 K/GPa. At low pressures $P\leq$0.49 GPa, the decrease of $T_c$ is accompanied  by both a weakening of 
the  spin correlations   at a finite wave vector and a reduction of   the density of states at the Fermi level. At high pressures above 1.93 GPa, however, the decrease of $T_c$ is due mainly to a reduction of   the density of states.   These results indicate that the electronic/magnetic state of Co is primary responsible for the superconductivity. Namely, both a large DOS and  spin correlations are indispensable for a high $T_c$ of this class of materials. The spin-lattice relaxation rate $1/T_1$ at $P$=0.49 GPa shows a $T^3$ variation below $T_c$ down to $T\sim 0.12T_c$, which provides compelling evidence for   the presence of  line nodes in the superconducting  gap function.  

We thank M. Nishiyama and K. Katayama for assitance in some of the measurements.This work was supported in part by research grants from MEXT, NSF, the T. L. L. Temple Foundation, the John and Rebecca Moores Endowment, and DoE.

\end{document}